\documentstyle[12pt]{article}

\begin{document}

\setlength{\baselineskip}{20pt}

\title{Calculation of the Density of States\\
Using Discrete Variable Representation and Toeplitz Matrices}
\author {Eli Eisenberg$^{1,2}$, Asher Baram$^1$ and Michael Baer$^1$\\ \\
$^1$Department of Physics and Applied Mathematics,\\ Soreq NRC, Yavne 81800,
Israel. \\ \\
$^2$Department of Physics, Bar-Ilan University, \\ Ramat-Gan 52900, Israel.}
\maketitle

\begin{abstract}
A direct and exact method for calculating the density of states for systems
with localized potentials is presented. The method is based on explicit
inversion of the operator $E-H$. The operator is written in the discrete
variable representation of the Hamiltonian, and the Toeplitz property
of the asymptotic part of the obtained {\it infinite} matrix is used. Thus,
the problem is reduced to the inversion of a {\it finite} matrix.
\end{abstract}

\newpage

The evaluation of the density of states has been
widely discussed for various physical systems in
condensed matter and chemical physics \cite{burke,rt,ho}.
In particular, one is interested in the exsistence of resonances,
and their position and width. In the field of chemical
physics, resonances play a role in electron-atom,
and atom-molecule scattering processes,
autoionization, associative detachment, dissociative atachment,
and similar molecular resonant reactions \cite{burke,tay1}.
Recently, the subject of resonant tunneling in semiconductors
double-barrier structures has also been
the subject of a feverish activity \cite{rt},
due to the technological interest of the properties of such
structures, e.g., negative differential resistance
and bistability in current-voltage response.

The common methods for finding the position of resonances are
the complex scaling method \cite{ho} and the stabilization method
\cite{tay1,tay2}. The complex scaling method, pioneered by the works
of Aguilar, Balsve and Combes \cite{abc}, and Simon \cite{bs} involves
the analytic continuation of the energy into the complex plane
using the transformation $r\to re^{i\theta}$, through diagonalization
of the scaled (non-Hermitian) Hamiltonian. The following complex
eigenvalues correspond to the resonant states, where the real part is
the resonance energy, and the imaginary part
is the life-time. The method was modified and extensively used
by Certain, Moiseyev and co-workers \cite{cer}, and has been
recently applied to three body problems \cite{lip}. However, this
method does not give the full density of states as a function of energy.
Only an approximated form can be obtained
through the sum of the Breit-Wigner functions of each resonance.

The stabilization method was developed in quantum chemistry
for problems involving electron-atom and electron-molecule scattering.
The basic idea is to repeatedly diagonalize the Hamiltonian in the basis
sets of ever larger extension $L$, from what is believed to be the region
where the resonance wave-function is localized. The result is a
stabilization diagram of the
eigenenergies $E_j(L)$ vs. $L$ (See an example on Figure 1).
The characteristic plateaus which contain the pattern of avoided
crossings between stable (with respect to $L$) and unstable eigenvalues,
where the former correspond to the eigenvalues representing resonances,
and the latter to discretized continuum states. Briefly, the explanation
of this method is the folowing \cite{mr}. The wavefunctions for energies
near the resonant energy are highly localized, and
thus are very well described by the finite $L^2$ states. Therefore, they
are stable with respect to changes in the range of the basis functions.
In contrast, eigenvalues far from resonance correspond to extended states,
and feel the modifications of the basis set. Thus, they will vary as $L$
changes.

Some methods have been suggested to derive the resonance parameters from the
stabilization method \cite{sim,bsd,mr2}. More recently, Mandelshtam {\it et
al} \cite{mand1} have shown how the full density of states can be obtained
using some kind of
averaging over the parameter $L$. This method has been used for dissociative
photoabsorbsion problems \cite{mand2}, for the calculation of
microcanonical and canonical rate constants for one-dimensional \cite{mand3}
and three-body collinear problems \cite{hel}, and for resonant
tunneling-times calculations \cite{porto}. A comparison of the above two
methods is given in Ref. \cite{ryb}

However, one would like to have a more direct way to calculate the density
of states, without applying analytic continuation methods, or
(somewhat artificial) averaging processes. For instance, the textbook
definition of the density of states is
\begin{equation}
\rho(E) = {\rm Tr}\delta(E-H) = -{1\over\pi}{\rm Im\ Tr}G(E),
\end{equation}
where $H$ is the Hamiltonian, and $G$ is the (full) outgoing Green's function
defined via $G(E)=(E+i\epsilon - H)^{-1}$. A clear and simple derivation
would be the evaluation of the Green's function through the inversion
of the matrix representing the operator $E-H$ in some $L^2$ basis set.
Unfortunately, this naive approach is not directly applicable, since the
physical systems in which resonant states occur are, in nature, of infinite
extent, and accordingly, the matrices involved are {\it infinite}. A
solution for this problem was given by Seideman and Miller \cite{sm} who
applied the method of negative imaginary potentials as absorbing boundary
conditions \cite{nip} to deal with the infinite asymptotes.

In this Letter, we present an alternative approach for the above problem.
We manage to invert the {\it infinite} matrix $E-H$ without applying to any
truncations (as done in the stabilization method) or imposing unphysical
boundary condintions. This is accomplished using the discrete variable
representation (DVR) \cite{olddvr,dvr,cm} in which the asyptotic parts are well
separated from the interaction region. It has been already
recognized \cite{jcp} that the asymptotic part of this matrix has
a Toeplitz structure \cite{toep}.
This structure is here used to reduce the problem into
the inversion of a finite matrix whose dimesion is proportional to the
width of the interaction region. The only parameter in this method is the
spacing $\sigma$ between successive grid points of the DVR.
We apply the method to two double barrier problem, with parameters set
corresponding to typical mesoscopic resonant tunneling and chemical
reaction problems respectively.

The essence of our method is employing the finite range of the potential
through the separation between the asymptotic regions and the interaction
region. In order to keep this separation, a localized basis set is desirable.
Apparantly, The most appropriate representation from this point of view is
the DVR, in which the potential operator is diagonal. The representation of
the kinetic energy part of the Hamiltonian is calculated through the infinite
order grid point representation of the second derivative, and is (for equally
spaced grid) of the form \cite{cm}
\begin{equation}
T_{ij}={\hbar^2 (-1)^{i-j}\over 2m\sigma^2}\left\{
\begin{array}{ll}
\pi^2/3 \hskip 1truecm& i=j \\
{2\over(i-j)^2} & i\neq j
\end{array}\right\delimiter0,
\end{equation}
where $m$ is the mass. The corresponding Green's function is thus
given through
\begin{equation}
(G^{-1})_{ij}= (E-V(x_i))\delta_{ij}-T_{ij},
\end{equation}
where the points $x_n=n\sigma$ are the basis grid points.
We use in the following only the outgoing Green's function,
and thus $E$ has to be understood as  $E+i\epsilon$.

We now look at the perturbative representation of $G$
\begin{equation}
\label{dyson}
G = G^0 + G^0VG^0 + G^0VG^0VG^0 + \cdots = SG^0,
\end{equation}
where
\begin{equation}
S = \sum_{n=0}^\infty (G^0V)^n
\end{equation}

In the following, we use a block-form in which the vectors (which correspond
to wave-functions) are represented by $3$-dimensional super-vectors whose
first (third) component correspond to the left (right) asymptotic part
of the vector (and is therefore an infinite-dimensional vector), and its
second component correspond to the interaction region (and is thus
finite-dimensional). Accordingly, the matrices are represented by $3\times 3$
super-matrices. In this notation, the potential- matrix can be written
as (we use bold letters for super-matrices elements, to stress that these are
matrices themselves)
\begin{equation}
V = \left(\begin{array}{ccc}
0&0&0\\
0&{\bf V}_{22}&0\\
0&0&0 \end{array}\right).
\end{equation}
Similarily, one can write the operator $G_0V$ as
\begin{equation}
G^0V = \left(\begin{array}{ccc}
0&{\bf G}^0_{12}{\bf V}_{22}&0\\
0&{\bf G}^0_{22}{\bf V}_{22}&0\\
0&{\bf G}^0_{32}{\bf V}_{22}&0\end{array}\right).
\end{equation}
Since this matrix has a zero (first and third) coloumns, so does
all its powers. One thus may write
\begin{equation}
S = \left(\begin{array}{ccc}
{\bf I}&{\bf A}&0\\
0&{\bf B+I}&0\\
0&{\bf C}&{\bf I}\end{array}\right),
\end{equation}
where the ${\bf I}$ operators are the identity operators of the
appropriate order for each block.

It is easy to see from the definition of $S$, that it satisfies the equation
\begin{equation}
S-I=G^0VS.
\end{equation}
Explicit multiplication, and comparing term by term, gives the following
relations
\begin{equation}
\label{sola}
{\bf A} = {\bf G^0}_{12}{\bf V}_{22}({\bf I+B}),
\end{equation}
\begin{equation}
{\bf B} = {\bf G^0}_{22}{\bf V}_{22}({\bf I+B}),
\end{equation}
\begin{equation}
\label{solc}
{\bf C} = {\bf G^0}_{32}{\bf V}_{22}({\bf I+B}).
\end{equation}
The solution of these relations is given by
\begin{equation}
{\bf B} = ({\bf I}-{\bf G^0}_{22}{\bf V}_{22})^{-1}
{\bf G^0}_{22}{\bf V}_{22},
\end{equation}
and consequently ${\bf A,C}$ are given by Eqs. (\ref{sola}),(\ref{solc}).
Note that the matrix ${\bf I}-{\bf G^0}_{22}{\bf V}_{22}$ is {\it finite}
and thus its inversion is a simple numerical problem.

Using (\ref{dyson}), the trace of $G$ is easily obtained. One has
\begin{equation}
\label{g}
{\rm Tr}(G-G^0) = {\rm Tr} ({\bf AG}^0_{21}) + {\rm Tr} ({\bf BG}^0_{22})
+ {\rm Tr} ({\bf CG}^0_{23}).
\end{equation}
The evaluation of the second trace involves simply a finite summation;
however, the two other traces are infinite sums. We will show now that
using the explicit form of $G^0$ these sums can be reduced to finite ones.

For this purpose we now calculate the DVR form of $G^0$. By definition,
\begin{equation}
G^0 = (E - T)^{-1},
\end{equation}
where $T$ stands for the kinetic energy operator. Using DVR, the matrix $E-T$
has the structure of a Toeplitz matrix, i.e., $(E-T)_{ij} = t_{i-j}$, where
\begin{equation}
t_n = \left\{\begin{array}{ll}
E - {\hbar^2\pi^2\over 6m\sigma^2}& n=0 \\
-{\hbar^2(-1)^n\over n^2m\sigma^2} & n\neq 0
\end{array}\right\delimiter0.
\end{equation}
The eigenvalues and eigenvectors are thus given by
\begin{equation}
\lambda^{(q)} = \sum_{n=-\infty}^\infty t_n e^{inq} =
E - {\hbar^2 q^2\over 2m\sigma^2};
\end{equation}
\begin{equation}
v^{(q)}_n = {1\over\sqrt{2\pi}}e^{iqn},
\end{equation}
where $q$ is a continuous index in the region $-\pi<q<\pi$ \cite{cac}.
The inverse matrix is thus given by
\begin{equation}
G^0_{ln} = \int_{-\pi}^\pi dq v^{(q)}_lv^{(q)*}_n{1\over\lambda^{(q)}} =
-{2m\sigma^2\over \pi\hbar^2}\int_0^\pi dq {\cos((l-n)q)\over \alpha^2
+ q^2},
\end{equation}
where $\alpha^2 = -(\sigma k)^2 -i\epsilon$, and $k$ is the wave number
corresponding to the energy $E$. The integrand is highly peaked around
$q=|\alpha|$, and therefore, whenever $|\alpha|<\pi$ and this peak is
inside the integration region, one can extend the integration region to
infinity, and obtain
\begin{equation}
\label{g0}
G^0_{ln} = -{im\sigma\over\hbar^2 k}e^{i\theta|l-n|},
\end{equation}
where $\theta = \sigma k$. The condition $\theta<\pi$ means that the
number of grid points per (free) wavelength is $> 2$. This is a relatively
sparse grid with respect to those which are used in usual DVR applications.

The free particle density of states per unit length is obtained from
(\ref{g0}),
\begin{equation}
\rho^0(E) = -{1\over\pi}{\rm Im\ Tr}G^0 = \sqrt{m\over 2E}
{1\over\pi\hbar}
\end{equation}

Substituting (\ref{g0}) in (\ref{g}), the infinite sums
reduce to geometric serieses, and one obtains the formula
\begin{equation}
\label{final}
\rho(E) = -{1\over\pi}\sum_{m,n = n_0+1}^{n_1-1} {\bf B}_{mn}
{\bf G^0}_{nm} + {\bf V}_m({\bf B+I})_{mn}\biggl(
e^{i\theta(m+n-2n_0)} + e^{-i\theta(m+n-2n_1)}
\biggr){(G^0_{00})^2\over 1-e^{2i\theta}},
\end{equation}
where $n_0$ ($n_1$) is the first point of the left (right) asymptotic region.
Thus, after inverting the matrix ${\bf B+I}$, all that one has to do is to
sum according to Eq. (\ref{final}).

As an example, we consider here a (symmetric) double barrier structure,
typical to the problems treated in mesoscopic resonant tunneling problems
\cite{porto}. The potential is of the form
\begin{equation}
\label{exam}
V(x)=V_0(\exp({\alpha(x-x_0)^2})+\exp({\alpha(x+x_0)^2}))
\end{equation}
The parameters used are $V_0 = 0.5eV$, $x_0 = 50 A$ and $\alpha = 4\times
10^{-3} A^{-2}$ (corresponding to a quantum well of width $\sim 60A$, with
barriers of width $\sim 40A$). The mass is $m^* = 0.041m_e$, corresponding
to the $\rm Ga_{0.47}In_{0.53}As$  well.
Figure 1 shows the full density of states, as calculated form Eq.
(\ref{final}). The resulting lineshape is a Lorenzian centered at $E_0 =
0.101180 eV$ whose half-width is $\Gamma/2=1.77127\times 10^{-3} eV$.
Figure 2 presents the same graph for the same potential
with set of parameters suitable for a chemical physics problem,
i.e., $V_0=0.5eV$, $x_0 = 0.2 A$, $\alpha = 50 A^{-2}$ and $m=m_p$.
The resulting lineshape is a Lorenzian centered at $E_0 = 0.3059235 eV$
whose half-width is $\Gamma/2=3.99496\times 10^{-3} eV$.

In conclusion, a direct and exact method for the calculation of the density
of states for localized-potential systems was derived. The method evaluates
the inverse of the matrix representing the operator $E-H$ in the DVR, employing
its asymptotic Toeplitz structure. As usual in DVR treatments, no integration
is needed in constructing the matrix elements. The numerical effort needed
involves only the inversion of one matrix (for each energy) whose
dimensionality is proportional to the range of the interaction region.
We have considered an explicit example of a one-dimensional double barrier
structure, typical to those considered in the field of resonant tunneling.
However, the derivation is completely general and can be applied also to
more complicated (e.g.,three-body, three dimensional) problems.

\noindent
---------------------------------------------

\bigskip\noindent
\newpage
{\bf \large Figure Captions:}

\bigskip\noindent
Fig. 1: The resonant part of the density of states, i.e., $\Delta\rho(E) =
	\rho(E)-\rho^0(E)$ for the potential (\ref{exam}). The full curve
	corresponds to our results, and the dotted one is the best Lorentzian
	fit.

\bigskip\noindent
Fig. 2: Same as Fig. 2 for second set of parameters.

\noindent
---------------------------------------------

\end {document}